\documentclass[11pt,twoside]{article}
\usepackage{asp2014}

\aspSuppressVolSlug
\resetcounters

\begin{document}

\title{Hybrid minimization algorithm for computationally expensive multi-dimensional fitting}

\author{Evgenii~Rubtsov$^{1,2}$, 
Igor~Chilingarian$^{3,1}$, 
Ivan~Katkov$^{1,4,5}$, 
Kirill~Grishin$^{1,6}$, 
Vladimir~Goradzhanov$^{1,2}$, and 
Sviatoslav~Borisov$^{7,1}$}

\affil{$^1$Sternberg Astronomical Institute, Moscow State University, Moscow, Russia; \email{evgenii.rubtsov@voxastro.org}}
\affil{$^2$Faculty of Physics, Moscow State University, Moscow, Russia}
\affil{$^3$Center for Astrophysics -- Harvard and Smithsonian, Cambridge, MA, USA}
\affil{$^4$New York University Abu Dhabi, Abu Dhabi, UAE}
\affil{$^5$Center for Astro, Particle, and Planetary Physics, NYU Abu Dhabi, Abu Dhabi, UAE}
\affil{$^6$Universit\'e de Paris, CNRS, Astroparticule et Cosmologie, F-75013 Paris, France}
\affil{$^7$Department of Astronomy, University of Geneva, Versoix, Switzerland}

\paperauthor{Evgenii Rubtsov}{evgenii.rubtsov@voxastro.org}{0000-0001-8427-0240}{Sternberg Astronomical Institute, Lomonosov Moscow State University}{}{Moscow}{}{119234}{Russia}
\paperauthor{Igor~Chilingarian}{igor.chilingarian@cfa.harvard.edu}{ORCID}{Center for Astrophysics -- Harvard and Smithsonian / Sternberg Astronomical Institute}{}{Cambridge}{MA}{02138}{USA}
\paperauthor{Ivan Katkov}{katkov.ivan@gmail.com}{0000-0002-6425-6879}{NYU Abu Dhabi}{Center for Astro, Particle, and Planetary Physics}{Abu Dhabi}{}{129188}{UAE}
\paperauthor{Kirill Grishin}{kirillg6@gmail.com}{0000-0003-3255-7340}{Sternberg Astronomical Institute, Lomonosov Moscow State University}{}{Moscow}{}{119234}{Russia}
\paperauthor{Vladimir Goradzhanov}{goradzhanov.vs17@physics.msu.ru}{0000-0002-2550-2520}{Sternberg Astronomical Institute, Lomonosov Moscow State University}{}{Moscow}{}{119234}{Russia}
\paperauthor{Sviatoslav Borisov}{sb.borisov@voxastro.org}{0000-0002-2516-9000}{University of Geneva}{Department of Astronomy}{Geneva}{}{}{Switzerland}



  
\begin{abstract}
Multi-dimensional optimization is widely used in virtually all areas of modern astrophysics. 
However, it is often too computationally expensive to evaluate a model on-the-fly.
Typically, it is solved by pre-computing a grid of models for a predetermined set of positions in the parameter space and then interpolating. 
Here we present a hybrid minimization approach based on the local quadratic approximation of the $\chi^2$ profile from a discrete set of models in a multidimensional parameter space. 
The main idea of our approach is to eliminate the interpolation of models from the process of finding the best-fitting solution. 
We present several examples of applications of our minimization technique to the analysis of stellar and extragalactic spectra.

\end{abstract}

\section{Introduction}

Successful interpretation of astrophysical phenomena in virtually all subfields of astrophysics are achieved by comparing observational data to numerical or analytic models and evaluating the physical parameters of these models.
Different minimization techniques are exploited to find a set parameters of such a model that  fits the input observing data best.
A large fraction of the minimization algorithms used in astronomy deal with continuous functionals in the parameter space.
However, evaluating the model on-the-fly for an arbitrary point in the parameter space often becomes too computational expensive (for example, to model a spectrum of star or a galaxy).
Typically, this problem is addressed by pre-computing a grid of models for a predetermined set of positions in the parameter space, which are then interpolated between grid nodes to achieve continuous and differentiable behavior of the evaluated function.
This approach, in cases with rapid change in the behavior of the models at neighboring nodes, leads to systematic artifacts caused by interpolation and an incorrect final solution.

Here we propose to eliminate the interpolation for models and search for an off-node solution using a local approximation of the $\chi^2$ profile by a positive definite quadratic form in a multidimensional parameter space.

\section{Minimization algorithm description}

\begin{figure}
    \centering
    \includegraphics[trim=100 50 100 50,clip,width=0.50\hsize]{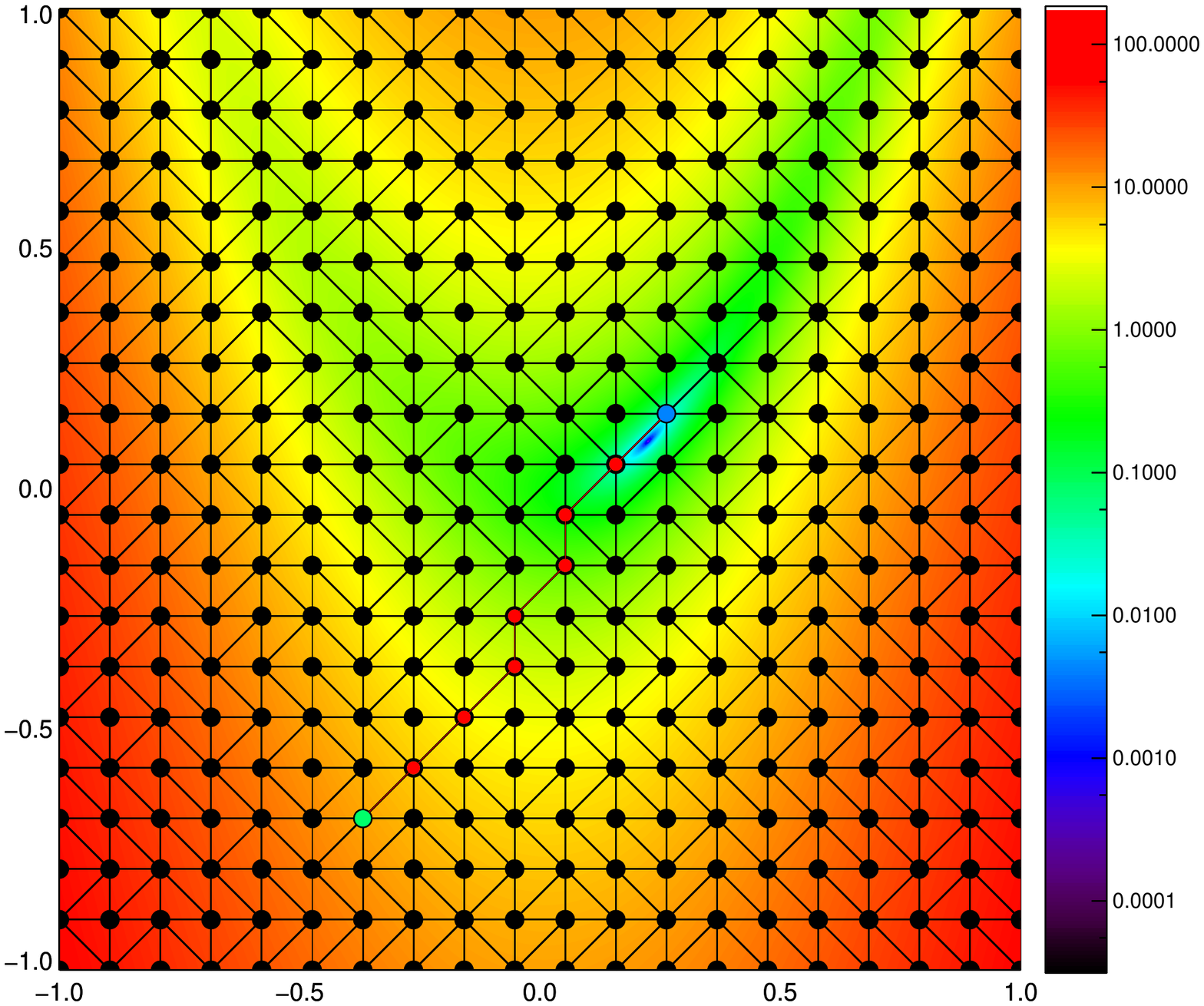}
    \hfill
    \includegraphics[trim=225 70 150 100,clip,width=0.39\hsize]{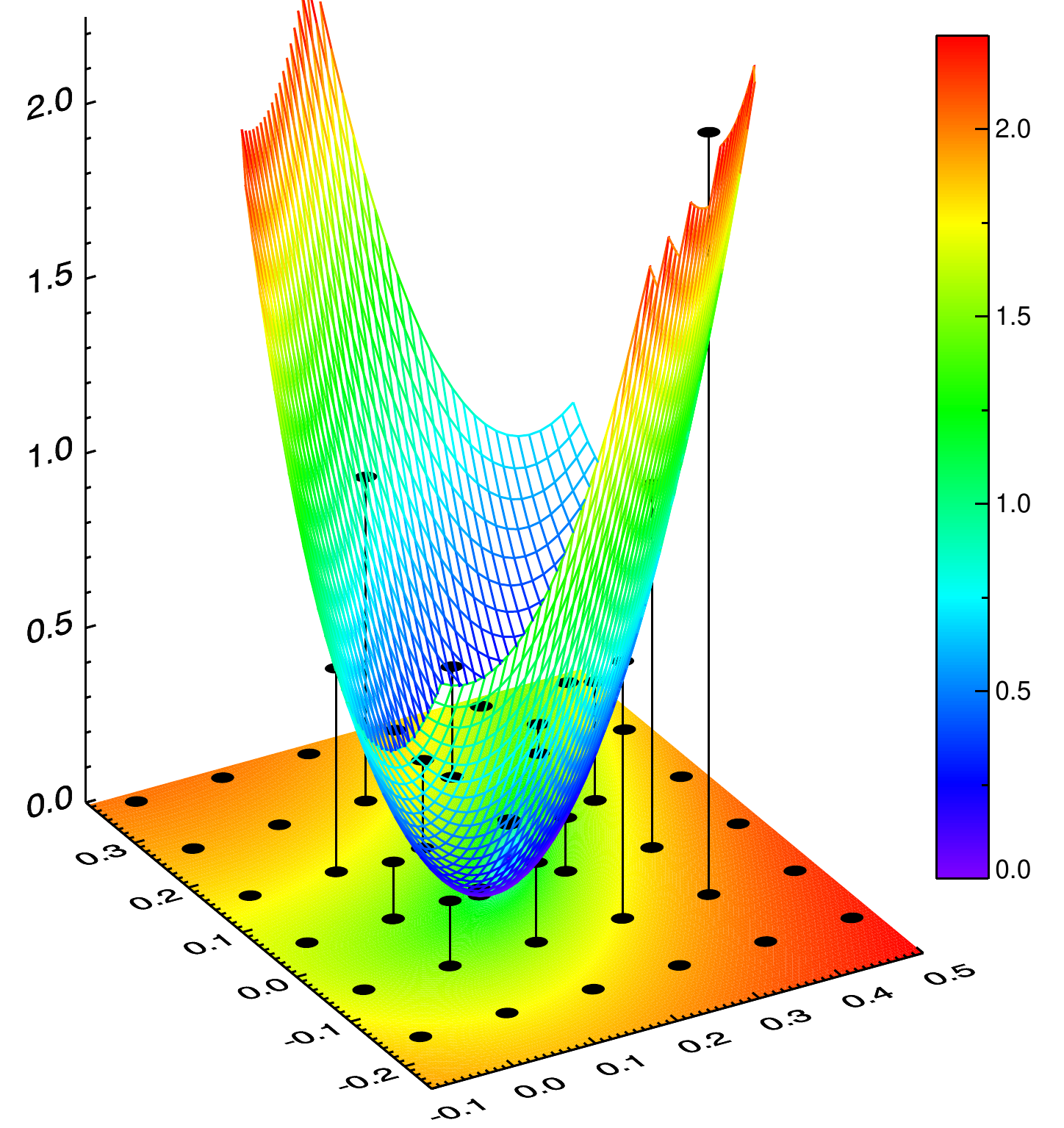}
    \caption{
An example of a synthetic test to validate our method for: a $\chi^2$ map uses the analytic Rosenbrock function (known for its notorious behavior in minimization tasks) with the minimum at $(0.225, 0.101)$.
The left panel shows the first part of the algorithm to locate the $\chi^2$ minimum grid node, a green point is the initial guess, red line connects the nodes with the detected minimum of $\chi^2$ on a grid, a blue point is the final node.
The right panel explains the second part of the algorithm with a local approximation of the $\chi^2$ profile and analytic determination of the minimum in the 2D parameter space $(0.234, 0.088)$.
        }
    \label{fig:fitting}
\end{figure}

The minimizer searches for the minimum of $\chi^2$ in the multidimensional parameter space defined by a discrete set of parameters (discrete parameters).
In our case, with the eliminated interpolation, the values of $\chi^2$ are determined only at the grid nodes.
Therefore, the minimization algorithm consists of two parts:
(i) searching the minimum value of $\chi^2$ at the grid nodes, which do not have to be regularly spaced;
(ii) calculating the off-node local $\chi^2$ minimum position in the parameter space.

The algorithm implementation includes: 
(i) construction a connectivity matrix for a grid of models using triangulation;
(ii) checking connected nodes and choosing a node with a minimum value of $\chi^2$ using a downhill/uphill climbing algorithm;
(iii) finding an off-node solution from the approximation of $\chi^2$ values at the locally connected nodes using a positively definite quadratic form \citep{Rosen2004};
(iv) determination of weights for the superposition of models at the minimum \citep{Saniee2008}.
This approach allows us to deal with irregular multidimensional grids of models, provided there is a local basis for of the corresponding dimensionality.

In addition to discrete parameters of the models, one can minimize continuous functional parameters (such as Doppler shift for radial velocities, rotational broadening for stars, velocity dispersions for galaxies, etc.) at each tested node using standard gradient methods (e.g. {\sc MPFIT}, \citealp{Markwardt2009}) (continuous parameters). Their final value can be obtained from a superposition of weights determined at the last step of the algorithm, or by supplying the final model into a ``standard'' minimizer to independently determine the values of continuous parameters. At this step we can also derive their statistical uncertainties, which are relevant if there is no template mismatch \citep{2020PASP..132f4503C}.

\section{Application of the algorithm}

\begin{figure}
    \centering
    \includegraphics[trim=35 192 60 205,clip,width=1.\textwidth]{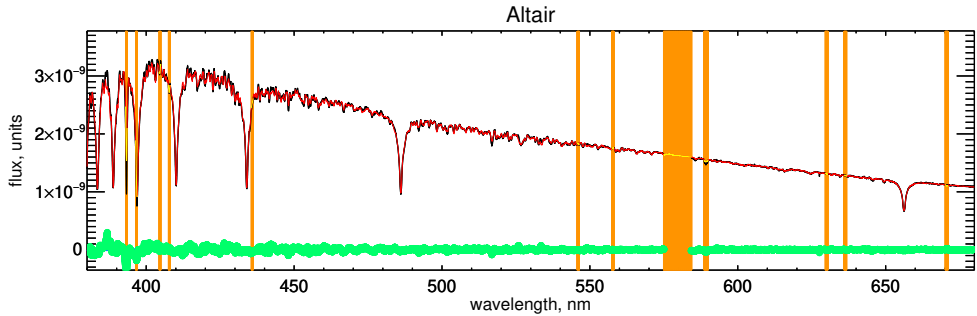}
    \caption{
An example of fitting the Altair spectrum from the UVES-POP stellar library using the new hybrid minimization, black, red, and green show observed fluxes, best-fitting models, and residuals; orange areas are excluded from the fit. 
    }
    \label{fig:uvespop_altair}
\end{figure}

\begin{figure}
    \centering
    \includegraphics[trim=35 192 60 205,clip,width=1.\textwidth]{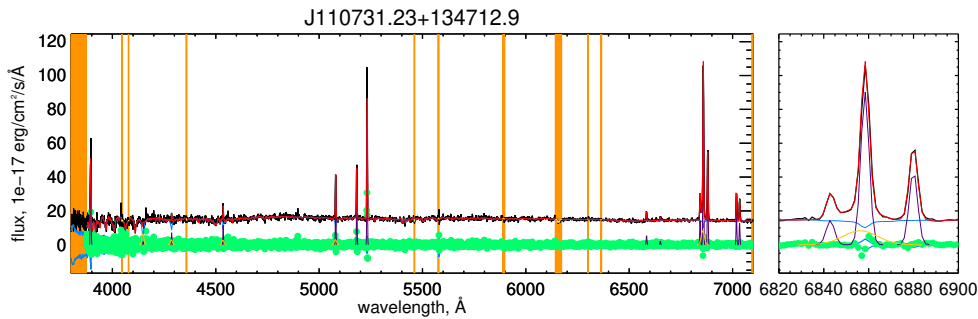}
    \caption{
An example of fitting an SDSS spectrum of the IMBH host galaxy \citep{Chilingarian+18} using {\sc NBursts} code with new minimization method. 
    }
    \label{fig:sdss_miles_mgfe}
    
    \vskip -0.7cm
\end{figure}

Before applying a new method to real data, we carried out a large number of synthetic tests on different test functions for optimization (artificial landscapes) with different grid configurations (regular/irregular).
The application of the method for the Rosenbrock function on a regular grid is shown in Fig.~\ref{fig:fitting}.

In the VOXAstro Stellar Libraries initiative (\url{https://sl-dev.voxastro.org/}), the new method showed excellent results consistent with the literature for UVES-POP  (\citet{Bagnulo2003} and Borisov et al. in prep), INDO-US \citep{Valdes2004} and ELODIE \citep{Prugniel2001, Prugniel2007} stellar spectra.
We also applied it to the new LCO-SL near-infrared stellar spectral library  \citep{Chilingarian2015}. In all these cases, we determined the following parameters of stellar atmospheres, $T_{\mathrm{eff}}$, ${\log g}$, ${\mathrm{[Fe/H]}}$, ${[\alpha/\mathrm{Fe}]}$ as discrete parameters for a synthetic grid of models computed with the {\sc phoenix} code \citep{Husser2013} and two continuous parameters ($v_{rad}$, $v_{rot}\sin{i}$).
An example of fitting result shown in Fig.~\ref{fig:uvespop_altair}.

As part of the RCSEDv2 (\url{https://rcsed2.voxastro.org/},  \citet{Chilingarian2017}) catalog project, we added the option to select a minimization algorithm to the {\sc NBursts} spectrum fitting technique \citep{Chilingarian2007a,Chilingarian2007b} and fitted large samples of galaxies with it.
Here, ages and metallicities of stellar populations were determined as discrete parameters using PEGASE.HR/MILES/E-MILES grids \citep{pegase,miles,e-miles}; radial velocities and velocity dispersions were set as continuous parameters.
An example of fitting result for a low-mass galaxy hosting an active galactic nucleus is shown in Fig.~\ref{fig:sdss_miles_mgfe}. We also plan to utilize this algorithm to estimate galaxy properties using multi-dimensional grids of stellar population models describing complex star formation histories \citep[see e.g.][]{2019arXiv190913460G}.

\section{Summary}

We developed a hybrid minimization algorithm, which allows us: (i) to deal with grids of models without interpolating them inside the evaluation routine; (ii) to deal with regular and irregular grids; (iii) to quickly build $\chi^2$ maps and look for a solution in different parts of the grid; (iv) to simultaneously determine discrete and continuous parameters of a model. The results of synthetic tests and examples of use in several research applications illustrate the power of using the hybrid minimization method in a wide range of astrophysical problems.

\acknowledgements This project is supported by the RScF grant 17-72-20119 and the Interdisciplinary Scientific and Educational School of Moscow University ``Fundamental and Applied Space Research''. ER is grateful to the ADASS-XXXI organizing committee for providing financial aid to support his attendance of the conference.

\bibliography{X7-003}


\end{document}